\documentclass[%
 reprint,
superscriptaddress,
 amsmath,amssymb,
 aps,
 prl
]{revtex4-2}

\usepackage{graphicx}
\usepackage{dcolumn}
\usepackage{bm}

\begin{document}

\preprint{APS/123-QED}

\title{Marangoni-driven flow instability accelerates liquid-solid contact on atomically smooth mica}

\author{Octave Castanedo}
 \affiliation{Université Paris-Saclay, ENS Paris-Saclay, DER de Physique, 91190, Gif-sur-Yvette, France.}
\affiliation{%
Institute of Mechanical Engineering, School of Engineering, \'{E}cole Polytechnique F\'{e}d\'{e}rale de Lausanne 
}%
\author{John M. Kolinski}%
 \email{john.kolinski@epfl.ch}
\affiliation{%
Institute of Mechanical Engineering, School of Engineering, \'{E}cole Polytechnique F\'{e}d\'{e}rale de Lausanne 
}%

\date{\today}

\begin{abstract}
A droplet falling toward a solid surface displaces the surrounding air until it encounters a defect, and contact nucleates. On atomically smooth surfaces devoid of defects, contact can be delayed until the droplet rebounds; however, above a critical impact velocity the droplet always contacts the surface. Here we show that for alcohol droplets in a humid atmosphere, the surface of the droplet above the solid develops patterns as a consequence of an instability; consequently, the liquid approaches the surface more closely in some places than others, accelerating contact formation. We demonstrate the attenuation and even suppression of this instability by varying the liquid composition, and characterize the growth rate and length scale of the patterns on liquid-air interface. 
\end{abstract}

\maketitle

When a liquid droplet falls toward a surface, it must drain the air beneath it before it can make contact. Depending on the impact velocity $V$, the droplet can trap a lubricating film of air; typically, this forms a dimple in the droplet's surface centered upon the impact axis\cite{thoroddsen_air_2003,  korobkin_trapping_2008, mandre_precursors_2009, hicks_air_2010, driscoll_ultrafast_2011, duchemin_curvature_2011, hicks_air_2012, bouwhuis_maximal_2012, hicks_liquidsolid_2013}; because this air film mediates the final moments before contact initiation, it plays a central role in governing the impact outcome. In the absence of surface roughness, a sub-micron-thick extended film of air remains entrained beneath the impacting droplet\cite{kolinski_skating_2012, de_ruiter_air_2015, li_probing_2015}. This film can cause the liquid to deviate up and away from the surface\cite{de_ruiter_dynamics_2012, kolinski_lift-off_2014}, and even completely rebound from the surface for $V<V_c$\cite{kolinski_drops_2014, de_ruiter_air_2015, kaviani_characteristic_2023, chubynsky_bouncing_2020, eggers_coalescence_2024, lewin-jones_collision_2024, sprittles_gas_2024}, where $V_c$ is a critical impact velocity; it can modify the dynamics of splashing droplets\cite{mandre_precursors_2009, mani_events_2010, mandre_mechanism_2012, kolinski_skating_2012, riboux_experiments_2014, li_probing_2015} or it can prevent \textit{elastomer} contact\cite{davis_elastohydrodynamic_1986, dillavou_virtual_2019, zheng_air_2021, bilotto_fluid-mediated_2024}. Fundamentally, the droplet impact dynamics - and the thin film of air beneath it - are highly sensitive to impact conditions\cite{yarin_drop_2006}. 

The liquid-gas interface beneath the droplet results in interfacial fluid flow - and such flows are often unstable\cite{craster_dynamics_2009}. Indeed, on scales near the capillary length, capillary stresses can drive fluid flows; for example, gradients in surface tension due to thermal-\cite{ehrhard_non-isothermal_1991, anderson_spreading_1995, oron_long-scale_1997, kalliadasis_marangoni_2003, sultan_evaporation_2005, dhaouadi_brethertons_2019} or concentration-gradients\cite{fanton_spreading_1998, pereira_dynamics_2008} lead to Marangoni-driven flows that are frequently unstable\cite{pearson_convection_1958, sternling_interfacial_1959, li_marangoni_2021}, driving a rich array of non-linear or unstable phenomena at interfaces\cite{stroock_fluidic_2003, keiser_marangoni_2017, kim_solutal_2017} and in droplets\cite{koldeweij_marangoni-driven_2019, lohse_physicochemical_2020, lopez_de_la_cruz_marangoni_2021, diddens_competing_2021, zeng_periodic_2021, zeng_droplet_2022, li_marangoni_2022, thayyil_raju_evaporation_2022}. While surface tension gradients emerge at the surface of droplets of alcohol in a humid atmosphere\cite{zhang_evaporation_1983, hu_analysis_2005}, and thermally driven fluxes are known to modify droplet impact outcomes for Leidenfrost droplets\cite{burton_geometry_2012, shirota_dynamic_2016, harvey_minimum_2021}, \emph{Marangoni-driven} flows have not been explored in the context of droplet impact processes and the thin film of air that forms during impact. Despite the abundant evidence that the thin film of air beneath an impacting droplet is sensitive to mechanical defects\cite{kolinski_drops_2014, kaviani_characteristic_2023}, the role of Marangoni flows during the impact of solvents, especially in a humid environment, is currently unknown. 

In this letter, we probe the sensitivity of the lubricating film of air beneath impacting droplets of isopropanol (IPA) on an atomically smooth, freshly cleaved mica substrate. The mica ensures that no mechanical defect is present beneath the impacting droplet; thus, we probe the ultimate stability of the lubricating film, as nucleation on the typically abundant mechanical defects is entirely suppressed. To probe the stability of the air film, we vary the environmental humidity and pressure, and the impact velocity, while directly imaging the liquid-air interface beneath the impacting droplet with Fizeau interferometry. We find a characteristic pattern emerges on the liquid-air interface, whose spatio-temporal dynamics are sensitive to the ambient humidity. By measuring the intensity contrast in the interferograms of the extended air film, we find that the pattern exhibits exponential growth; furthermore, the spatial dependence of the pattern, as measured about an azimuth centered on the impact axis, defines a characteristic wavelength. Using this spatio-temporal information, we measure the growth rate of the surface disturbance as a function of the wavelength, and find that this is sensitive to the ambient conditions, consistent with a Marangoni-driven instability. Finally, we show that by modifying the purity of the solvent, and probing the impact of solvent-water mixtures, we can suppress the instability. An experiment conducted in dry nitrogen suggests that the observed instability is driven by solutal, rather than thermal, Marangoni effects.

\begin{figure}
\includegraphics[width=\linewidth]{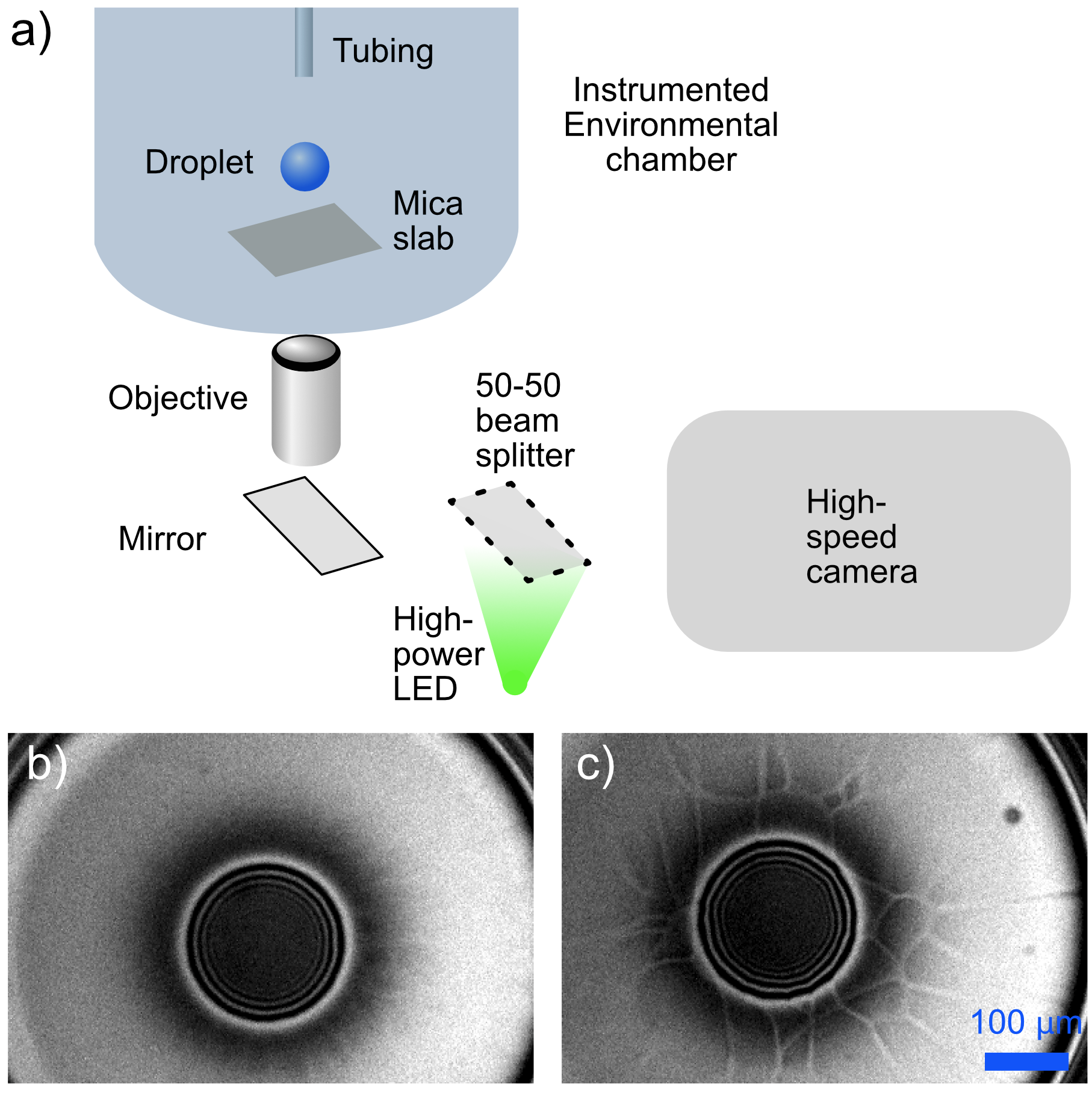}
\caption{Experimental schematic and typical images. a) A high speed camera is attached to an interference microscope that images the surface beneath the impacting droplet onto the camera's sensor. Reflections from the solid surface and the droplet-air interface generate an interference pattern consisting of peaks of intensity and troughs of intensity as integer multiples of a gap $h = \lambda/4$. The droplet is released from a thin Teflon tube and made to impact on a freshly cleaved mica surface. The experiment can is carried out in an instrumented environmental chamber, where the ambient pressure and relative humidity are monitored. b) a typical image with elevated contrast for low relative humidity show a predominantly symmetric air film, whereas c) for high relative humidity, a distinctive pattern breaks the air film's azimuthal symmetry.    }
\label{fig1}
\end{figure}

Droplets of 99\% pure isopropanol with $< 0.1$ wt~\% rhodamine 6G dye are generated by applying pressure to a syringe. The syringe drives the fluid through a teflon fluid line to a luer adapter (diameter 0.55 mm) that is suspended above the impact surface on a three-axis precision displacement stage. The fluid line serves as a junction between the inside and outside of a PMMA environmental chamber that is used both to control the humidity and the gas composition in the volume around the droplet impact experiment. On the bottom surface of the experimental chamber is a glass window that enables direct observation of the impact surface from below. Freshly cleaved mica surfaces are coupled to the window with immersion fluid (Nikon Type-N). Images of the underside of the drop are recorded with a high-speed camera (Photron Nova S16) connected to a long-working distance microscope with an in-line illumination for the interferometry. The droplet impact apparatus and optical configuration are depicted schematically in Fig.~\ref{fig1} a). 


During the impact of droplets of water, silicone oil or alcohol in low-ambient humidity, the liquid-air interface remains smooth at long times\cite{kolinski_drops_2014, kaviani_characteristic_2023}. This smoothness appears as an intensity field that is symmetric about the impact axis. A typical image of these dynamics, recorded for an alcohol droplet in a low-relative humidity ambient approximately 500 $\mu$sec after the droplet entered the field of view, is shown in Fig.~\ref{fig1} b). For alcohol droplets in a high-humidity ambient environment, however, the interfacial structure is fundamentally different. Symmetry about the impact axis is broken, and a pattern emerges on the underside of the droplet corresponding to changes in air film thickness; a typical image of these dynamics is shown in Fig.~\ref{fig1} c), also 500 $\mu$sec after the droplet entered the field of view. 

\begin{figure}
\includegraphics[width=\linewidth]{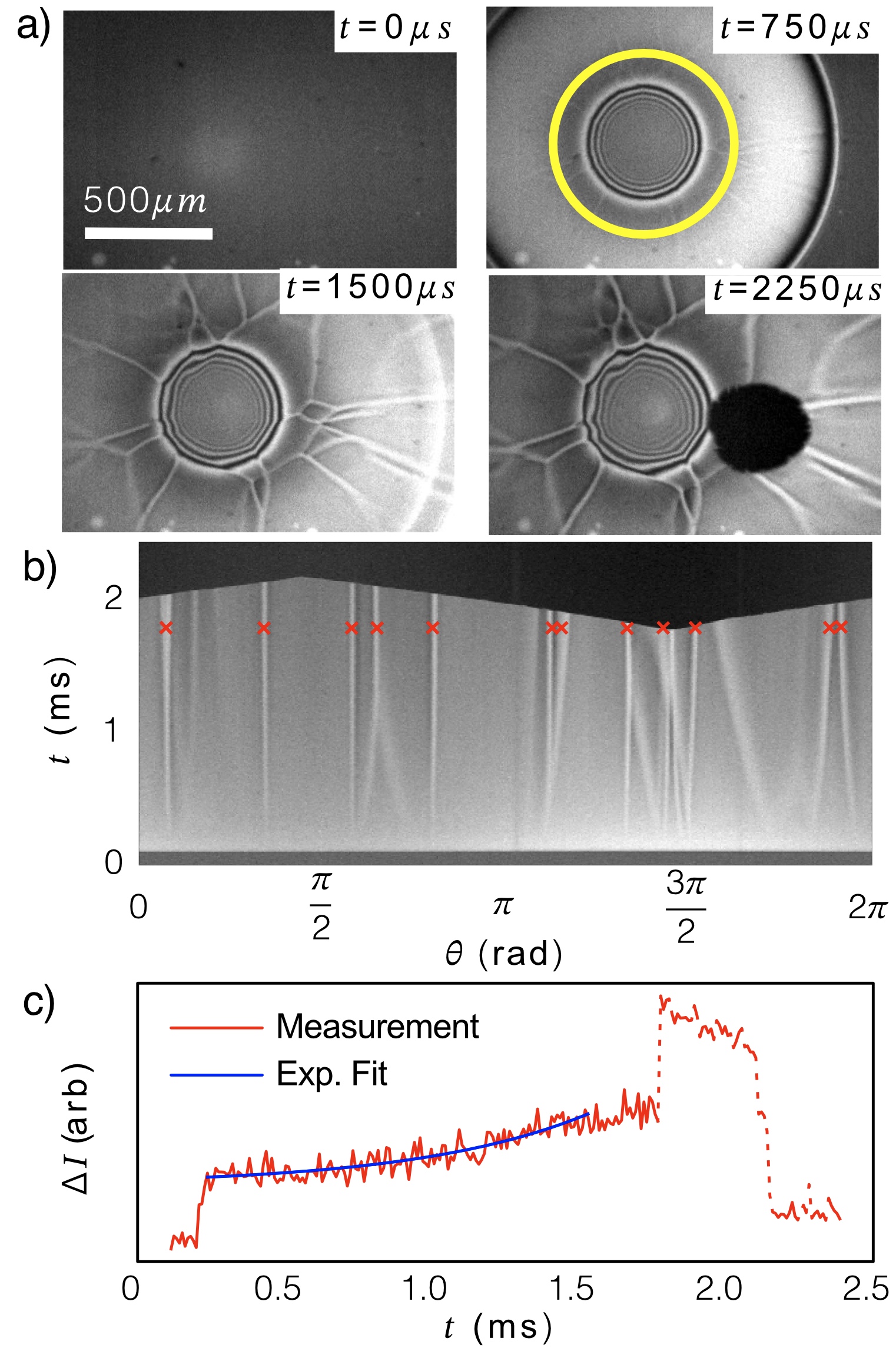}
\caption{Temporal evolution of the surface pattern. a) A time-series of the evolving interface beneath the droplet shows that the intially smooth surface develops a more pronounced pattern over time. b) A kymograph of the intensity sampled about the center of the impact axis and unfolded azimuthally shows the growth of the pattern. Red `x's indicate the peaks in the pattern, the number of which, $N$, is used to define a wavenumber $k$ to parameterize the spatial structure of the pattern about the impact axis. c) The growth rate and the length-scale of the pattern can be readily extracted from this kymograph, as shown for the example developed from analyzing the time series shown in (a). The dashed line corresponds to data after contact formation.   }
\label{fig2}
\end{figure}

The pattern that emerges beneath impacting solvent droplets at elevated ambient humidity evolves over time, as can be seen in the time series in Fig.~\ref{fig2} a). As is readily apparent in the time series, the structure of the instability changes, and the depth of intensity gradients becomes greater. To characterize these dynamics, we define an azimuth at radius $r_A = 420\mu$m centered about the impact axis, as illustrated by the yellow circle in the 2nd frame of the time series in Fig.~\ref{fig2} a). The intensity measured along this azimuth is unwrapped about angle $\theta$, and plotted as a function of time in a kymograph, as shown in Fig.~\ref{fig2} b). This graph encodes the spatio-temporal dynamics as symmetry breaks about the impact axis. As is readily apparent, the intensity variation 
grows as time increases. From the data in the kymograph, we extract the amplitude of the interface by measuring the difference between the maximum and minimum intensity along the intensity trace, defining an intensity span at time $t$ as $\Delta I (t) = (I_{max}- I_{min}) (t)$. Here, we have labeled the local maxima, as seen in Fig.~\ref{fig2} b).

Prior to saturation, the intensity span $\Delta I$ initially grows with an exponential character as a function of time $t$, as can be seen in the graph of $\Delta I (t)$ in Fig.~\ref{fig2} c). The growth rate is extracted by an exponential fit to these data using a non-linear least squares routine implemented in Python\cite{virtanen_scipy_2020} ; in selecting the fitting range, we avoid the earliest stage of the dynamics where the liquid enters the depth of field of the microscope, and truncate the fitting range when the signal saturates, as indicated by a strong change of slope. 

To characterize the spatial scale of the symmetry breaking about the impact axis, we count the number of peaks $N$ about the azimuth (those indicated by the red `x' symbols in Fig.~\ref{fig2} b)) and divide by the azimuthal distance, $2 \pi r_A$. This defines an effective wavenumber $k$ for the instability as $k = \frac{N}{2 \pi r_A}$.

Droplets impacting on atomically smooth mica surfaces typically rebound below a threshold impact velocity\cite{kolinski_drops_2014}, $V_{cr}$. $V_{cr}$ is highly sensitive to the ultimate stability of the air film\cite{kaviani_characteristic_2023}, and thus serves as a watershed in the droplet impact phase diagram between contact or not. We exploit this sensitivity of $V_{cr}$ to explore the effect of the observed instability on the formation of contact in the absence of mechanical nucleation on a defect by measuring $V_{cr}$\footnote{The impact velcoity $V$ is determined by precisely measuring the droplet release height $H$, and calculating the ballistic velocity upon impact as $\sqrt{2 g H}$.} as a function of the ambient pressure in the chamber, $p$, for both the IPA and water-glycerol mixtures with a viscosity of approximately 10 cSt as a control. Notably, the liquid air interface beneath water-glycerol solution droplets does not destabilize in a fashion similar to that observed in Fig.~\ref{fig2} a)\cite{kaviani_characteristic_2023}. The majority of experiments are carried out at relative humidity, $RH>50$ \%. As a control, we also identified the critical threshold velocity for rebound at a lower ambient humidity value of $RH \sim 35$ \% and $p_{atm}$. 

\begin{figure}
\includegraphics[width=0.7\linewidth]{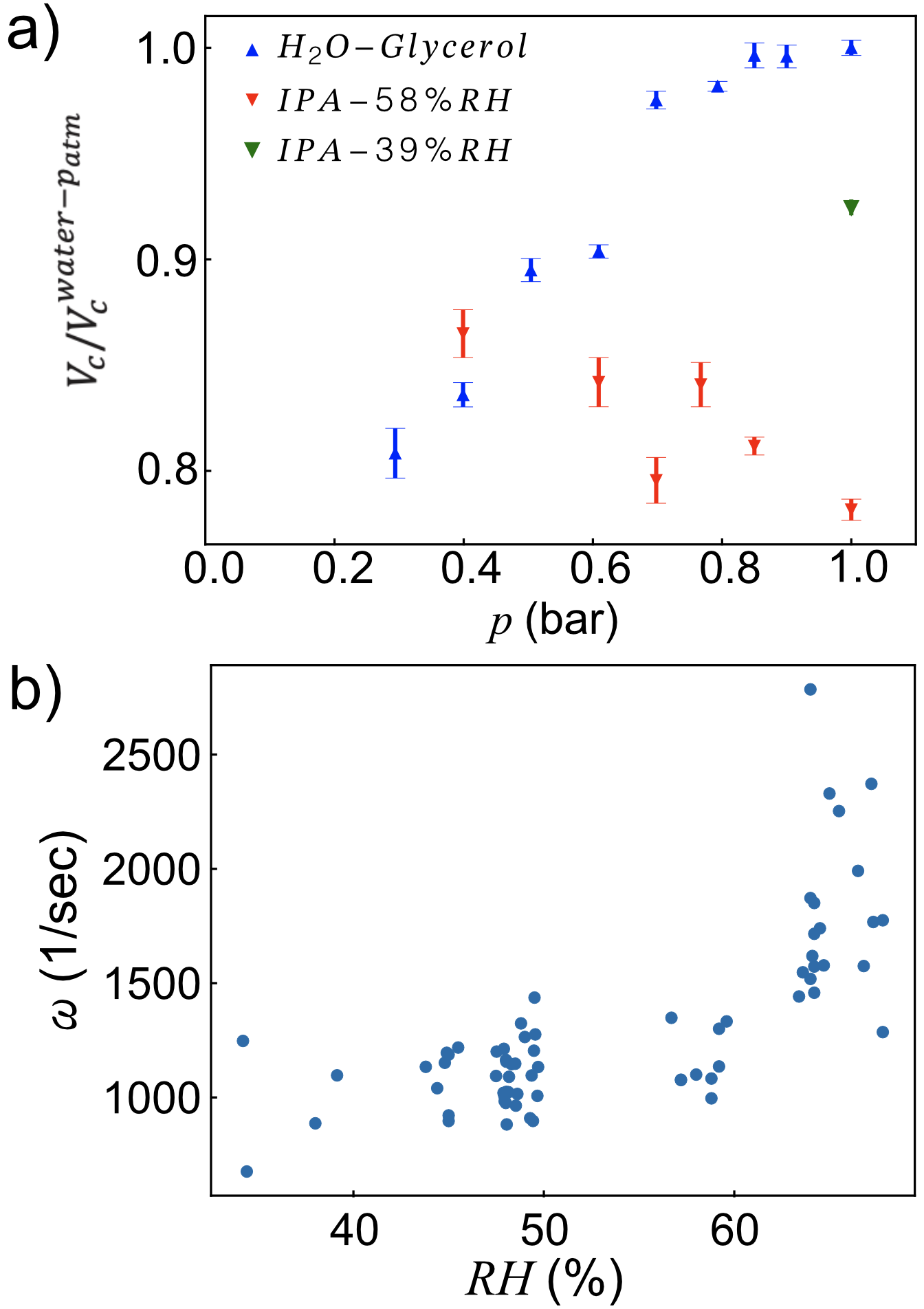}
\caption{Acceleration of contact formation for reduced pressure under high humidity conditions. a) Baseline experiments with water-glycerol solution show a monotonic trend of the critical impact velocity for rebound as a function of ambient pressure. For alcohol under high ambient humidity conditions (RH = 56.5 $\pm$ 2.5\%), the formation of contact is more rapid at atmospheric pressure than at reduced pressure, as can be seen by the rebound-contact threshold. This trend persists down to pressures as low as 0.4 atm. For alcohol under lower humidity, the threshold velocity at atmospheric pressure is higher than under higher humidity. 
b) Analysis of the growth rate for the surface instability at $p_{atm}$ shows that the liquid-air interface is more unstable at higher humidity conditions for a given pressure than at lower humidity conditions. }
\label{fig3}
\end{figure}

The threshold velocity for rebound of the water-glycerol droplets decreases as the ambient pressure is reduced, at first gradually, and then more steeply as the pressure is further decreased, as can be seen in Fig.~\ref{fig3} a). The nonlinear response of the critical velocity as $p$ is reduced suggests that the air film becomes less stable as the ambient pressure decreases. Surprisingly, at high ambient humidity conditions, the \textit{opposite} trend is observed for the IPA drops; indeed, the air film becomes \textit{more stable} at reduced pressure. By contrast, at \textit{lower} RH, corresponding to a dryer ambient, and atmospheric pressure, the alcohol drop has a \textit{higher} critical velocity for rebound than at any pressure or RH, indicating a more stable liquid-air interface. These trends can be seen in the graph of relative critical velocity as a function of $p$ in Fig.~\ref{fig3} a). 

To explore the humidity dependence of the instability at the liquid-air interface, we measured the growth rate for a large number of experiments while varying the ambient RH at atmospheric pressure; we find that the instability grows much more rapidly as humidity increases beyond 60 \%, and that the growth rate increases as a strongly non-linear function of RH, as can be seen in Fig.~\ref{fig3} b). While the growth rate increases slowly for 30\% $<$ RH $<$ 60\%, it is nevertheless increasing; the hallmark pattern of the instability is clearly visible throughout this range, with slight modifications at lower RH.  

\begin{figure}
    \includegraphics[width=\linewidth]{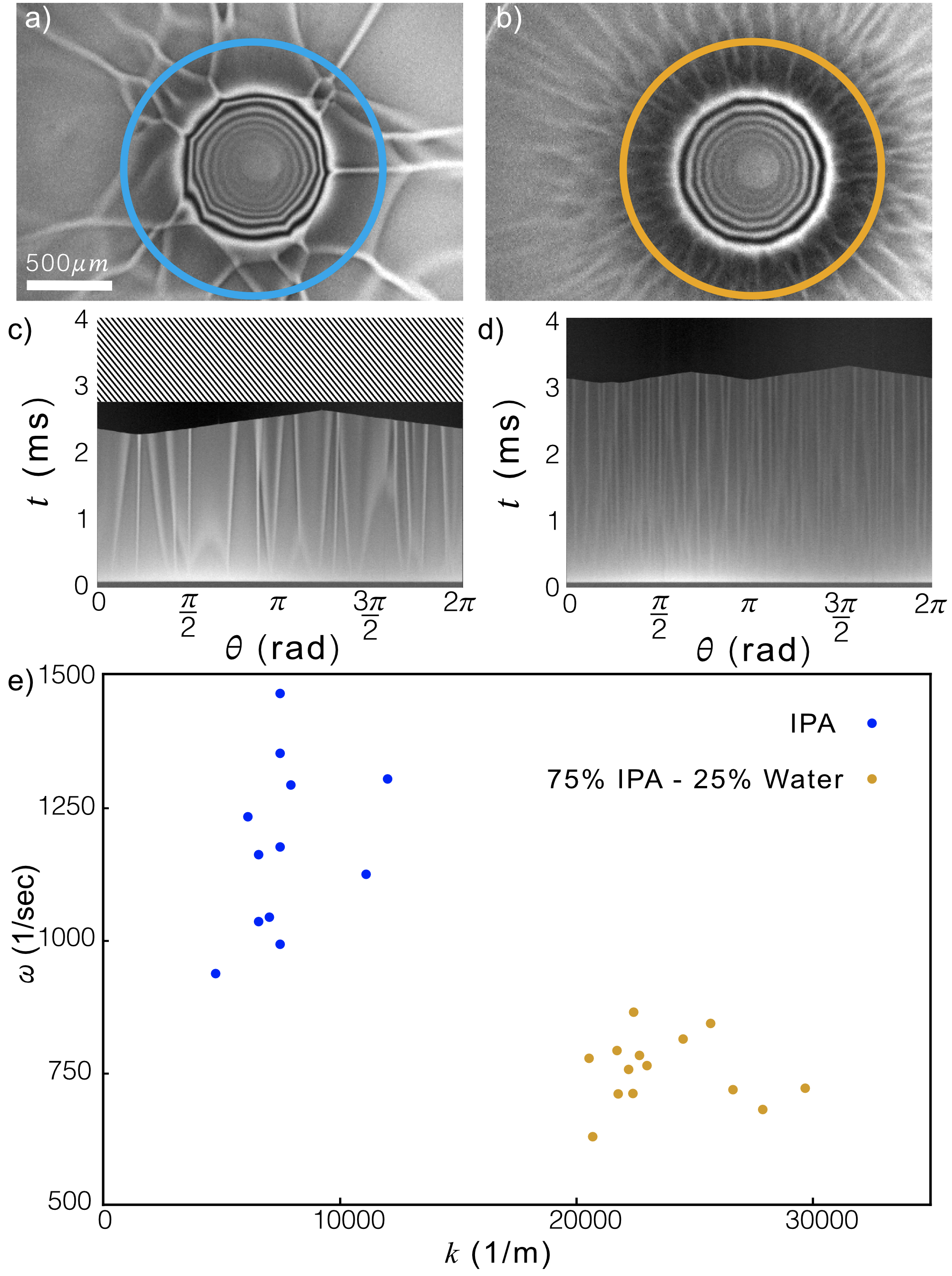}
    \caption{Attenuation of the instability at elevated water concentration. a) A comparison of the instabilities for droplets of different liquids (mixture 75\% IPA and 25\% water on top and pure IPA at the bottom). b) A comparison of the kymographs corresponding to the images on a). c) The growth rate  of the surface pattern for droplets impacting at $V = 0.577$ m/sec is graphed as a function of the wavenumber for neat IPA and the IPA-water mixture. The growth rate is significantly less for the elevated water concentrations and the wavenumber is significantly higher, resulting in a higher amplitude perturbation at a given time, with greater spacing between features.}
    \label{fig4}
\end{figure}

To probe the sensitivity of the spatio-temporal dynamics of the emergent pattern in the air film during impact, we altered the concentration of water in the droplet source liquid by dilution for several humidity levels from RH = 40 \% to RH = 58\%. To this end, we mixed 3 parts alcohol with 1 part water. At this level of dilution, a drastically different pattern emerges: both the amplitude of the pattern, represented by the intensity contrast within the imaged air film region, and the spacing of the patterned features is significantly smaller than for the neat IPA droplet, with identical ambient humidity. Two typical images are shown for the neat IPA and the mixture in Fig.~\ref{fig4} a) \& b). 

The spatio-temporal dynamics of the pattern seen in the snapshots in Fig.~\ref{fig4} a) \& b), can be seen in the corresponding kymographs shown in Fig.~\ref{fig4} c) \& d), as unwrapped around an azimuth of the impact axis along the colored circles at $r_A = 420 \mu$m. As is apparent from the snapshots, the wavenumber is greater, and the growth rate less, for the solution than for the neat IPA. This suggests that the instability leads to a weaker amplitude perturbation of the interface\footnote{Linear stability analysis requires an initial amplitude to translate to physical growth. Here, a relevant amplitude scale for initial perturbations of the interface might be the thermo-capillary length, $\ell_{tc} = \sqrt{\frac{k_B T}{\gamma}}$, where $k_B T$ is the thermal energy, and $\gamma$ is the surface tension. For water at room temperature, $\ell_{tc} = 4.5$ Angstroms, which would not be visible in the initial images, but would roughly correspond to an amplitude of a few nm in one growth time, slightly below estimated amplitude at the measured growth rates.} over the timescale of the experiment for this liquid, just as is observed at low RH for neat IPA.

To evaluate whether the trend apparent from the snapshots and kymographs in Fig.~\ref{fig4} a)-d) is consistent across many experiments, we carry out a similar analysis to that of Figs~\ref{fig2} \&~\ref{fig3} to measure the growth rate and dominant wavenumber for each impact experiment. These data are graphed across a large number of experiments for neat IPA and the mixture, as shown in Fig.~\ref{fig4} e). The growth rate is greater, and at a lesser wavenumber, for the neat IPA, whereas the growthrate is less, and at a larger wavenumber, for the mixture. This suggests that the air film is more stable for the mixture than for the neat IPA, consistent with the observations of the snapshots and the kymographs in Fig.~\ref{fig4} a)-d).

Taken as a whole, the observations here suggest that the nanometric air film beneath an impacting droplet of IPA can be unstable to Marangoni stresses that are generated when the alcohol droplet absorbs water from the ambient environment. Marangoni stresses generate extremely complex dynamics at a fluid-fluid interface\cite{pearson_convection_1958, lohse_physicochemical_2020}, and can be driven by two different mechanisms - either solutal, where the absorbed water locally modifies the surface tension, driving a flow at the interface; or thermal, where gradients in evaporative flux locally modify the surface tension by thermal gradients. 

To test whether solutal or thermal Marangoni stresses dominate during droplet impacts, we further modified the ambient conditions by replacing the humid lab air with dry nitrogen gas. We found that the appearance of the instability of the IPA-air film profile was completely suppressed upon realizing this impact condition - this suggests with high likelihood that the observed instability of the liquid-air interface is \emph{solutal}, rather than \emph{thermal}; if the evaporative flux were to generate thermal gradients and associated Marangoni-driven flows, these would also appear in the absence of ambient humidity, but the experiment in a dry nitrogen atmosphere shows no evidence of this. 

While our measurements identify the stability of the interface at a consistent value of radius from the impact center, the spatio-temporal dynamics are in fact quite rich even away from this region. Some key questions motivated by the image include, what determines the persistence of the spatial structure of the instability as the liquid front deforms during the formation of the air film? Why does the amplitude and wavelength decrease with increasing distance from the impact center? Indeed, the observation that the spacing and amplitude decrease as the droplet spreads outward may indicate some type of coursening behavior. Furthermore, it is unclear whether the formation of the instability at the initial stage of the impact, which generates channels, can feed back on the flow in the air film to modify the subsequent dynamics.

The characterization carried out to analyze the stability of the interface, where we identify the growth of $\Delta I$ and measure a wavenumber $k$ associated to this growth by using the peaks in the azimuthal kymograph as plotted in Fig.~\ref{fig2} b) is motivated by a typical linear stability analysis. Though the peaks we identify are clearly in a saturated regime, the persistence of the pattern over time and the exponential character of the observed growth suggests that the linear-stability analysis provides insight into the dynamics of the interface when it becomes unstable; nevertheless, contact formation appears to be a scale-separated instability\cite{kaviani_characteristic_2023} when the solutal Marangoni instability may already be in saturation, based on the separation of timescales for the growth of the interfacial pattern and the initiation of a contact patch during impact. 

The theory for solutal Marangoni instability at a liquid-liquid interface is rich and very subtle\cite{pearson_convection_1958, sternling_interfacial_1959}. As a function of the governing parameters, several stability boundaries emerge in the interfacial dynamics. One central question concerning the applicability of existing stability analysis is how the boundary conditions modify the growth rate and wavelength of solutal Marangoni instability beneath the impacting droplet. Indeed, the boundary condition on the mica substrate, and the thickness of the air film, should significantly modify the state of stress in the air film, which should scale with the ratio of the air velocity to the air film thickness. Here, the air film is below 1 $\mu$m in thickness, and thus the stress may be very high even though the gas viscosity is small. Because the air film is so thin, transport of mass and momentum are highly impeded; the rate at which the gas chemistry and volume equilibrate at the onset of the instability will be much slower than a free gas volume. Perhaps this enhances persistence of water concentration gradients in the droplet that drive the instability.

Given the ubiquity of the phenomenology of droplet impact and the vast number of studies on droplet impact dynamics, including those focused on the lubricating film of air, it is worth considering why this instability has not been reported in prior literature. A possible explanation for this is the typical formation of contact nucleii long before the instability has a chance to appreciably grow; essentially, on glass or other surfaces, one does not readily maintain an intact air film of sufficient duration to observe droplet rebound\cite{kolinski_drops_2014, de_ruiter_air_2015}, and thus only the earliest instants of the formation of the instability might be visible. Indeed, the reliability of freshly cleaved mica to generate atomically smooth surfaces is instrumental in realizing the conditions where the solutal Marangoni instability can develop sufficiently to be readily observed with interferometry\cite{driscoll_ultrafast_2011} or frustrated Total Internal Reflection methods\cite{kolinski_skating_2012, shirota_measuring_2017}, or combinations thereof\cite{kaviani_high_2023}.

The ultimate formation of contact is a localized rupture of the air film. The air film rupture dynamics are driven by interfacial attraction of the liquid solution to the mica surface, which has an attractive interaction due to van der Waals forces. An intriguing question that remains open following our study is whether the same van der Waals forces that drive contact formation \emph{locally} may also modify the growth rate and contribute to enhancing the solutal Marangoni driven instability we observe here. 

We have shown that contact formation can be accelerated on atomically smooth mica surfaces (the ultimate contact regime) for solvents falling through humid air. This contact formation rate is accelerated by an instability driven by solutal Marangoni flows that perturb the smoothness of the interface, and drive some regions of the liquid into closer proximity to the mica surface than others, essentially breaking the symmetry of the liquid-air interface. These observations highlight the importance of a potentially new regime of interfacial instability with a strong confinement of one of the fluid phases, with implications for industrial processes including manufacturing and heat transfer. While we have identified the instability, and characterizes its growth rate and spatial scales, significant open questions remain - for instance, does the confinement of the gas film modify the stability threshold or the dynamics? And to which extent does the linear stability analysis predict contact formation, which seems to emerge after the linear instability might have already reached a non-linear saturation? And which mechanisms are potentially responsible for non-linear saturation in this setting? Further studies are required to address these questions.

\begin{acknowledgments}
We wish to acknowledge Tobias Schneider and Detlef Lohse for helpful comments and suggestions.
\end{acknowledgments}

%

\end{document}